\documentclass[amsmath, amssymb, aps, prb, superscriptaddress, floatfix, reprint,journal=jctc]{revtex4-2}

\usepackage{booktabs}  
\usepackage{array}     
\usepackage{paralist}  
\usepackage{verbatim}  
\usepackage{subfigure} 
\usepackage{mathrsfs}
\usepackage{amsmath}
\usepackage[version=4]{mhchem}
\usepackage{float}
\usepackage{dsfont}
\usepackage[table]{xcolor}
\usepackage[titletoc]{appendix}
\usepackage{xtab}
\usepackage{longtable}
\usepackage[utf8]{inputenc}

\usepackage{braket} 
\usepackage{subfigure} %
\usepackage{graphicx}
\usepackage{makecell} 

\usepackage{placeins} 
\usepackage{epstopdf}
\usepackage{tikz}
\usepackage[hidelinks]{hyperref}
\usepackage{xurl}

\graphicspath{{Azomethane/Figures/}}

\begin{document}
\title{
Nonadiabatic excited-state dynamics with quantum Monte Carlo-trained machine learning: azomethane as a stringent test
}

\author{Alfonso Annarelli}
\author{Emiel Slootman}
\author{Claudia Filippi}
\email{c.filippi@utwente.nl}
\affiliation{MESA\textsuperscript{+} Institute for Nanotechnology, University of Twente,  7500 AE Enschede, The Netherlands}
\date{\today}

\begin{abstract}
We introduce quantum Monte Carlo (QMC)-trained multi-state machine-learned (ML) force fields for nonadiabatic excited-state dynamics, targeting photochemical processes in which the electronic character changes along the reaction path and a consistent correlated description is required. In this framework, variational Monte Carlo wave functions combine compact selected configuration-interaction expansions with a Jastrow factor that explicitly accounts for dynamical correlation, while neural networks convert the stochastic QMC data into smooth potential energy surfaces for large surface-hopping ensembles. We apply this approach to azomethane, a demanding test case involving torsional relaxation through conical-intersection regions and C--N bond dissociation on the hot ground state. Benchmark calculations support the accuracy of the QMC reference data and show robust force convergence across isomerization and dissociation geometries.
The QMC-trained dynamics preserves the expected photoisomerization mechanism, strongly reduces the excessive C--N breaking obtained with complete active space self-consistent field, and predicts a small but non-negligible prompt dissociation component after internal conversion, with a timescale consistent with femtosecond-resolved mass-spectrometry experiments. These results establish QMC-ML as a practical route to nonadiabatic photochemical dynamics with accurate wave-function reference data.
\end{abstract}

\maketitle

\section{Introduction}

Nonadiabatic molecular dynamics simulations of photochemical processes require an accurate and balanced description of potential energy surfaces (PESs) across multiple electronic states, particularly in regions of strong nonadiabatic coupling such as conical intersections \cite{curchod_review,Worth_review,nonadiabatic_review}.
Recent studies \cite{Gonzalez_electrstruct,Janos_electstruct,Papineau,janos_2026} have shown that the outcome of ultrafast nonadiabatic dynamics can depend more strongly on the underlying electronic structure method than on the details of the nuclear-electronic propagation scheme. This places stringent requirements on the electronic structure level used to generate the trajectories.

Multireference methods such as complete active space self-consistent field (CASSCF) \cite{Roos_casscf,Roos_casscf87} and (extended) multi-state CAS second-order perturbation theory (CASPT2) \cite{Wolinski_caspt2,Finley_mscaspt2,xms} are standard choices for such simulations, although single-reference approaches can be adequate in specific contexts \cite{Olivucci_adc2,Barbatti_singleref,Zhu_tddft}.
A central challenge is to maintain a balanced treatment of the relevant electronic configurations and of dynamical correlation across all regions sampled by the trajectories. In methods based on a fixed active space, this balance depends critically on the orbitals and configurations selected \textit{a priori}. Important configurations may become missing as the electronic character changes or bonds rearrange, leading to geometry-dependent errors in potential energy surfaces and forces. Changes in orbital character can also produce convergence problems and crashed trajectories \cite{Barbatti_fullycorrelated,Papineau}. Perturbative corrections improve the treatment of dynamical correlation but may introduce additional numerical difficulties near degeneracies or in the presence of intruder states \cite{intruder_states}.

Here, we use variational Monte Carlo (VMC) combined with configuration interaction using a perturbative selection made iteratively (CIPSI) \cite{Huron_cipsi} to generate reference energies and forces for multi-state machine-learned (ML) nonadiabatic photodynamics. Building on previous applications of VMC/CIPSI to excited states \cite{Dash_excited_cipsi, PinedaFlores2019a,Dash2021,Cuzzocrea_2022,Shepard_2022}, 
the Jastrow-Slater wave functions combine compact and flexible determinantal expansions with an explicit treatment of dynamical electron correlation. All wave function parameters are optimized state-specifically, with orthogonality to lower-lying states enforced through penalty constraints \cite{Pathak_2021}. By targeting a common perturbative correction in the generation of the determinantal expansions, the approach provides a practical route to comparable wave-function quality across molecular geometries and electronic states, without relying on a manually selected active space.

Direct VMC dynamics over large trajectory ensembles remains impractical because of both computational cost and stochastic noise in the forces. We therefore use neural-network force fields \cite{Unke_mlff} as a smooth surrogates for the VMC/CIPSI reference data. While quantum Monte Carlo (QMC) methods have already been used to construct ML force fields for ground-state molecular and condensed-phase systems \cite{Tirelli2022,Niu_2023,Huang_dmcml,Huang2023a,Tenti2024,Slootman_forces}, their use as reference data for multi-state machine-learned nonadiabatic photodynamics has not yet been explored.

\begin{figure}[htbp]
\begin{center}
\includegraphics[width=8.6cm]{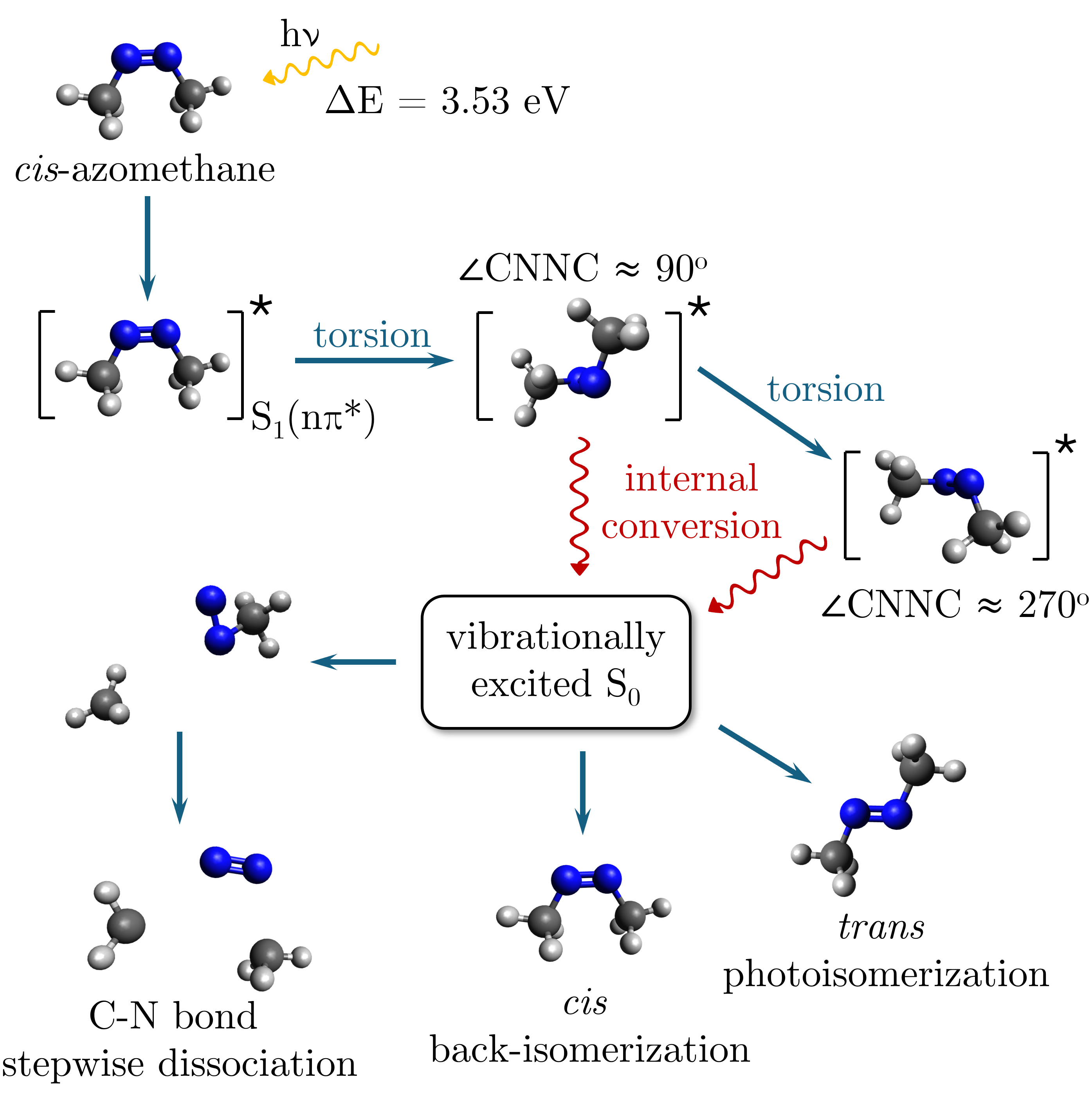}
\caption{Schematic representation of gas-phase \textit{cis}-azomethane photodynamics, showing excitation to the lowest excited state $S_1$, torsion of the $\angle$CNNC dihedral angle, internal conversion to the hot ground state $S_0$, and subsequent isomerization and dissociation channels.}
\label{fig: reaction scheme}
\end{center}
\end{figure}

\begin{sloppypar}
We demonstrate this machine-learned QMC approach on azomethane (\ce{CH3-N=N-CH3}), a prototypical molecule displaying a rich interplay of photo-induced ultrafast processes \cite{cattaneo_persico_azoinsolution,Sellner_photodyn,Sellner_azoinsol,Diau_concert,Diau_femtochem,North_expt}. Upon excitation to the lowest singlet excited state, \textit{cis}-azomethane undergoes torsional relaxation towards conical-intersection regions and internal conversion to the hot ground state, where it can recover the \textit{cis} form, isomerize to \textit{trans}, or undergo C–N bond dissociation as illustrated in Figure~\ref{fig: reaction scheme}.
We first assess the VMC/CIPSI reference data through vertical excitation energies and force comparisons along relaxation and dissociation pathways. These benchmarks probe regions with significantly different electronic character and establish the robustness of the QMC description across the configuration space relevant to the dynamics. We then train multi-state neural-network force fields on the VMC/CIPSI energies and forces and perform large-ensemble surface-hopping simulations, comparing the resulting photodynamics with analogous CASSCF- and CASPT2-based models. Finally, by extending the dynamics to \textit{trans}-azomethane, we revisit the long-standing debate on the timescale of azomethane dissociation and assess whether early C--N bond cleavage is an artifact of the electronic structure description or a genuine post-internal-conversion component, comparing with previous simulations and femtosecond-resolved experiments. 
\end{sloppypar}

\section{Methods}

\subsection{Variational Monte Carlo Forces}


Here, we employ variational Monte Carlo (VMC) \cite{Foulkes2001,Luchow_review,Lester_review} to compute expectation values of observables, cast as high-dimensional integrals over electronic coordinates. In particular, the energy is computed as
\begin{equation}
    E=\int E_L(\mathbf{R})P(\mathbf{R})\mathrm{d}\mathbf{R}\equiv \langle E_L\rangle_P,
\end{equation}
where $\mathbf{R}$ denotes the $3N$ electronic coordinates. The quantities $E_L(\mathbf{R})= \hat{H}\Psi(\mathbf{R})/\Psi(\mathbf{R})$ and $P(\mathbf{R})=|\Psi(\mathbf{R})|^2/\int |\Psi(\mathbf{R'})|^2\mathrm{d}\mathbf{R'}$ are the local energy associated with the trial wave function $\Psi$ and the probability distribution sampled in a VMC run, respectively. The energy is then estimated by averaging the local energy over a set of electronic configurations distributed as $P(\mathbf{R})$.

Similarly, the atomic forces can be rewritten as an average of a local force estimator over the distribution $P(\mathbf{R})$ \cite{Reynolds_forces,Assaraf2003,Filippi_derivatives,Badinski_2010}:
\begin{equation}
    F=-\nabla_\alpha E=-\langle \nabla_\alpha E_L(\mathbf{R}) + (E_L(\mathbf{R})-E)\nabla_\alpha \ln P(\mathbf{R}) \rangle_P.
\end{equation}
While this force estimator obeys a zero-variance principle in the limit that the wave function and its derivatives are exact, it displays an infinite variance for an approximate trial function. To address this issue, we adopt a guiding wave function that differs from the trial function close to the nodes and remains finite at the nodes \cite{Sorella_cutoff}. As a measure of the distance from the nodes, we use $d = |\nabla \Phi / \Phi|$, where $\Phi$ is the determinantal component of the wave function. 

\subsection{Trial Wave Function}

The trial wave function used here is of the so-called Jastrow-Slater form:
\begin{equation}
    \Psi(\mathbf{R})=\mathcal{J}(\mathbf{R})\sum_i c_i\, D_i(\mathbf{R}),
\end{equation}
where $D_i$ are Slater determinants built from single-particle orbitals.
The Jastrow factor, $\mathcal{J}$, depends explicitly on the interelectronic distances and is expressed as the exponential of polynomials in the electron-electron and electron-nucleus distances, capturing both two-body (electron–electron, electron–nucleus) and three-body (electron–electron–nucleus) correlation contributions. Unless otherwise stated,  all calculations are performed with a two-body Jastrow factor. 

The starting determinantal component of the Jastrow-Slater wave function is generated from CIPSI expansions~\cite{Huron_cipsi}.
In this selected-CI approach, determinants are added iteratively according to their second-order perturbation (PT2) energy contribution. For multi-state calculations, a common determinantal space is built by selecting determinants according to the weighted average over the states of their PT2 contributions. The weights are adjusted during the selection so that the states have comparable total PT2 energy corrections, which we use as a measure of the quality of the wave function. The expansions are grown until a target total PT2 energy correction is reached, and the same target correction is used for all azomethane geometries and electronic states. The impact of using determinantal components based on CASSCF calculations is also tested in Section~S3.


All wave function parameters (Jastrow, CI, and orbital coefficients) are optimized by energy minimization for all states of interest, and orthogonality between a given excited state and all lower-lying states is enforced through a penalty-based state-specific approach \cite{Wheeler_2024,Pathak_2021,Shepard_2025},
in which the following state-specific functional is minimized
\begin{equation} \label{eq: state-specific energy}
    E_{SS}[\Psi_I] = E_I[\Psi_I] + \sum_{J<I} \lambda_{IJ} \frac{\braket{\Psi_I|\Psi_J}^2}{\braket{\Psi_I|\Psi_I}\braket{\Psi_J|\Psi_J}},
\end{equation}
where the coefficients $\lambda_{IJ}$ determine the strength of the constraint and are chosen sufficiently large relative to the relevant energy separations between states.

\section{Computational Details}

The QMC calculations are performed with the CHAMP code \cite{filippi_champ} using scalar-relativistic energy-consistent Hartree-Fock pseudopotentials and the corresponding correlation-consistent Gaussian basis sets \cite{BFD_pseudo,DolgFilippi_private}.
Unless otherwise stated, we use the aug-cc-pVDZ basis set. Increasing the basis to aug-cc-pVTZ yields statistically compatible excitation energies at the \textit{cis} equilibrium geometry (Table~S2) and changes the forces at the selected benchmark configurations by at most 0.8~kcal/mol/\AA, as measured by the mean absolute deviation (MAD) between the two basis sets (Figure~S4).

The determinantal components of the Jastrow-Slater wave functions are obtained from CIPSI calculations with Quantum Package \cite{Quantum_package}. Natural orbitals are generated from preliminary CIPSI runs starting from state-averaged (SA)-CASSCF(2,2) orbitals over the two lowest states, computed with GAMESS(US) \cite{GAMESS}. The final expansions are built in these natural orbitals using a target PT2 energy correction of $-0.616$ a.u.\ for all geometries and states. The convergence with respect to the number of selected determinants is discussed in the SI. The interface of both Quantum Package and GAMESS(US) with CHAMP uses the TREXIO library \cite{Posenitskiy2023}. A two-body Jastrow factor \cite{jastrow_2body} is used throughout. Tests on selected configurations show that adding electron-electron-nucleus terms to the Jastrow factor changes the forces by at most $1.2$~kcal/mol/\AA\ in MAD (Figure~S4).
The wave functions are optimized with the stochastic reconfiguration method \cite{Sorella_stoch_reconf,Neuscamman_opt}, using the state-specific functional of Eq.~\ref{eq: state-specific energy} with $\lambda_{10}=1$ a.u. Forces are computed with a node-cutoff parameter $\epsilon=0.1$ a.u.\ \cite{Sorella_cutoff}.


All-electron coupled cluster with single and double excitations and perturbative triples [CCSD(T)] calculations are performed with Psi4 \cite{Psi4} using the cc-pVQZ basis set. 
All-electron SA(2)-CASSCF, multi-state (MS)-CASPT2, and extended XMS-CASPT2 calculations over the two lowest states are carried out with OpenMolcas \cite{Openmolcas}, using the aug-cc-pVDZ and aug-cc-pVTZ basis sets. 
For (X)MS-CASPT2, we employ the standard IPEA shift of 0.25 a.u.~\cite{Ghigo2004a} and an imaginary level shift~\cite{Forsberg1997a} of $0.1$ a.u. The active spaces considered are CAS(2,2), with two active electrons in the nitrogen lone-pair $n_-$ and $\pi^*$ orbitals; CAS(6,4), obtained by adding the occupied $n_+$ and $\pi$ orbitals; and CAS(12,10), which further includes the relevant C--N and N--N $\sigma$ and
$\sigma^*$ orbitals. In the following, SA(2)-CASSCF is abbreviated as CASSCF.

\textit{Ab initio} nonadiabatic molecular dynamics simulations are performed with curvature-driven \cite{BaeckAn,doCasal_Barbatti} trajectory surface hopping  ($\kappa$TSH) \cite{Tully_surface_hopping,Barbatti_surface_hopping}, as implemented in SHARC 4.0~\cite{Mai2018_SHARC} and interfaced with OpenMolcas.
The curvature-driven approximation avoids the explicit evaluation of nonadiabatic couplings or wave function overlaps \cite{Hammes_Schiffer_Tully} between different geometries, while giving comparable results for this photoisomerization process \cite{Zhao_ktdc}. This is further supported by our CASSCF(12,10) and MS-CASPT2(12,10) tests (Figure~S12), where $\kappa$TSH and overlap-based excited-state populations agree.

The nuclear and electronic time steps are 0.5 and 0.02 fs, respectively, and each \textit{ab initio} trajectory is propagated for 150 fs. We use the energy-based decoherence correction with a decay factor of 0.1 a.u.\ \cite{Granucci_decoherence}. After a successful hop, velocities are rescaled along their directions to conserve the total energy, while no adjustment is applied after frustrated hops. This choice is consistent with previous studies showing that different rescaling schemes have negligible effects on the dynamics \cite{Barbatti_velocadjust}.
The initial conditions are sampled from the Wigner distribution of the vibrational ground state in the harmonic approximation \cite{DahlSpringborg_1988}.
The \textit{cis} and \textit{trans} equilibrium geometries are optimized at the B3LYP/cc-pVTZ level with GAMESS(US) and the corresponding harmonic frequencies are computed at the same level with OpenMolcas.

The adaptive sampling \cite{Behler_2015,Behler_2017,Westermayr_longml} procedure used to generate the ML training configurations and the subsequent model generation are carried out with the SPaiNN package~\cite{spainn,Painn_architecture} interfaced with SHARC~4.0. The configurations are generated once, in an adaptive-sampling workflow at the CASSCF level. Starting from CASSCF(6,4), an initial set of 752 configurations is collected. The active space is then enlarged to CASSCF(12,10), allowing the adaptive sampling to explore additional regions of the PESs, in particular those involving C--N dissociation. The sampling is continued to convergence yielding a final dataset of 2320 configurations.  Energies and forces for CASSCF(12,10), (X)MS-CASPT2(12,10), and QMC are computed on this set of configurations to train the corresponding ML models. The sampling protocol, hyperparameters, prediction errors on the test set, and dynamical checks against reference \textit{ab initio} trajectories are reported in the SI.

ML–driven surface hopping trajectories are propagated for 400 fs using the same settings as the \textit{ab initio} simulations. A total of 1000 trajectories is used to ensure statistical convergence of the excited-state populations and allow an estimation of the associated uncertainties.

\section{Results}

The lowest singlet excited state of azomethane has predominantly $n \rightarrow \pi^*$ character. Population of this state weakens the N--N bond and initiates torsional motion of the $\angle$CNNC dihedral angle toward the region of strong nonadiabatic coupling.
After internal conversion, the excess energy deposited on the ground-state surface can also promote C--N bond cleavage.

The dissociation mechanism has long been debated \cite{Burton_expt,North_expt,Fairbrother_expt,Bracker_expt,Gejo_expt} and the most recent experimental studies using femtosecond-resolved mass spectrometry \cite{Diau_concert,Diau_femtochem} support the stepwise process,
\begin{equation}
\ce{C2H6N2 -> N2CH3^{\bullet} + CH3^{\bullet} -> N2 + 2CH3^{\bullet}},
\end{equation}
but it remains unclear whether the first methyl group dissociates within tens of femtoseconds after crossing the conical intersection (``impulsive" model) or  on longer, picosecond timescales, after equilibration on the ground-state potential energy surface (``statistical" model).

Although \textit{trans}-azomethane is the most stable configuration \cite{West_azom,hutton_azom}, we focus primarily on the photoisomerization from \textit{cis}-azomethane because it has been extensively studied computationally \cite{Sellner_photodyn,Papineau,cattaneo_persico_azoinsolution} and displays richer dynamics, with two passages through conical intersections and a higher dissociation probability. The \textit{trans}-initiated dynamics is then used as a
complementary test case: it probes the same torsional relaxation and C--N bond cleavage after internal conversion in a simpler setting, with a more direct connection to the ultrafast photofragmentation experiments.

The results are organized as follows. We first assess the electronic structure description through vertical excitation
energies and force comparisons along the relevant relaxation and dissociation pathways. We then compare the \textit{cis}- and \textit{trans}-initiated dynamics obtained with ML force fields trained on QMC, CASSCF, and MS-CASPT2 reference data, with additional XMS-CASPT2 results reported in the SI.

\subsection{Vertical excitation energies}

Table~\ref{tab: vertical energies} reports the lowest vertical excitation energies of \textit{cis}- and \textit{trans}-azomethane. The VMC/CIPSI values, which are well converged with respect to the CIPSI expansion size (see SI), are compared with extrapolated full CI (exFCI) estimates, CASSCF and MS-CASPT2 results, and available literature and experimental data.

The VMC excitation energies closely reproduce the exFCI results for both isomers. They also fall in the range of available high-level literature values such as multireference configuration interaction (MRCI) and CC3.
The CASSCF excitation energies increase with the size of the active space, while MS-CASPT2 shifts the values downward and underestimates the VMC and exFCI results. The close agreement of VMC with exFCI indicates that the CIPSI-based trial wave functions provide a balanced description of the lowest vertical excitation in both isomers.  

\begin{table}[ht]
\centering
\caption{Vertical excitation energies (eV) of \textit{cis}- and \textit{trans}-azomethane at the optimal B3LYP/cc-pVTZ  geometries. CASSCF and MS-CASPT2 calculations use the aug-cc-pVTZ basis set, while VMC and exFCI use aug-cc-pVDZ. The statistical error and extrapolation uncertainty on the last digit are indicated in parenthesis.}
\begin{tabular}{l l l }
\hline \addlinespace
Method & $\Delta$E$_{\rm \textit{cis}}$ & $\Delta$E$_{\rm \textit{trans}}$ \\ \hline \addlinespace
CASSCF(6,4)    & 3.65 & 3.84 \\
CASSCF(12,10)  & 3.98 & 4.17 \\
MS-CASPT2(6,4)    & 3.20 & 3.43 \\
MS-CASPT2(12,10)  & 3.36 & 3.57 \\
VMC/CIPSI     & 3.581(4) & 3.769(4) \\
exFCI        & 3.53(1) & 3.75(1) \\
ADC(2)$^a$        & 3.50 & 3.66 \\
MRCI$^b$          & 3.62 & 3.82 \\
CC3$^c$           &      & 3.76 \\
Expt.\ (gas phase) \cite{Robin_expt_gap}      &      & 3.65 \\
\addlinespace \hline
\multicolumn{3}{p{6cm}}{$^a$RI-ADC(2)/aug-cc-pVTZ from Ref.~\citenum{Sellner_photodyn}.\newline $^b$MRCI(6,4)/6-31G$^*$ from Ref.~\citenum{Sellner_photodyn}.\newline $^c$CC3/aug-cc-pVDZ from Ref.~\citenum{SZALAY}.}
\end{tabular}
\label{tab: vertical energies}
\end{table}

\subsection{Force benchmarks along isomerization and dissociation pathways}

To test the VMC/CIPSI forces, we select a representative set of twelve azomethane geometries relevant to the dynamics. Seven configurations are taken along an excited-state minimum energy path from the \textit{cis} geometry toward the conical intersection, which involves initial bond relaxation followed by torsion of the $\angle\text{CNNC}$ dihedral angle toward $90^\circ$. Four additional configurations are selected from a surface-hopping trajectory exhibiting double C--N dissociation in the ground state, and one corresponds to single C--N dissociation leading to the metastable $\ce{N2CH3^{\bullet}}$ intermediate \cite{cattaneo_persico_azoinsolution}. This set therefore probes the two main regions of interest, namely, isomerization and C--N dissociation.

\begin{figure}[bhtp]
\begin{center}
\includegraphics[width=7.4cm]{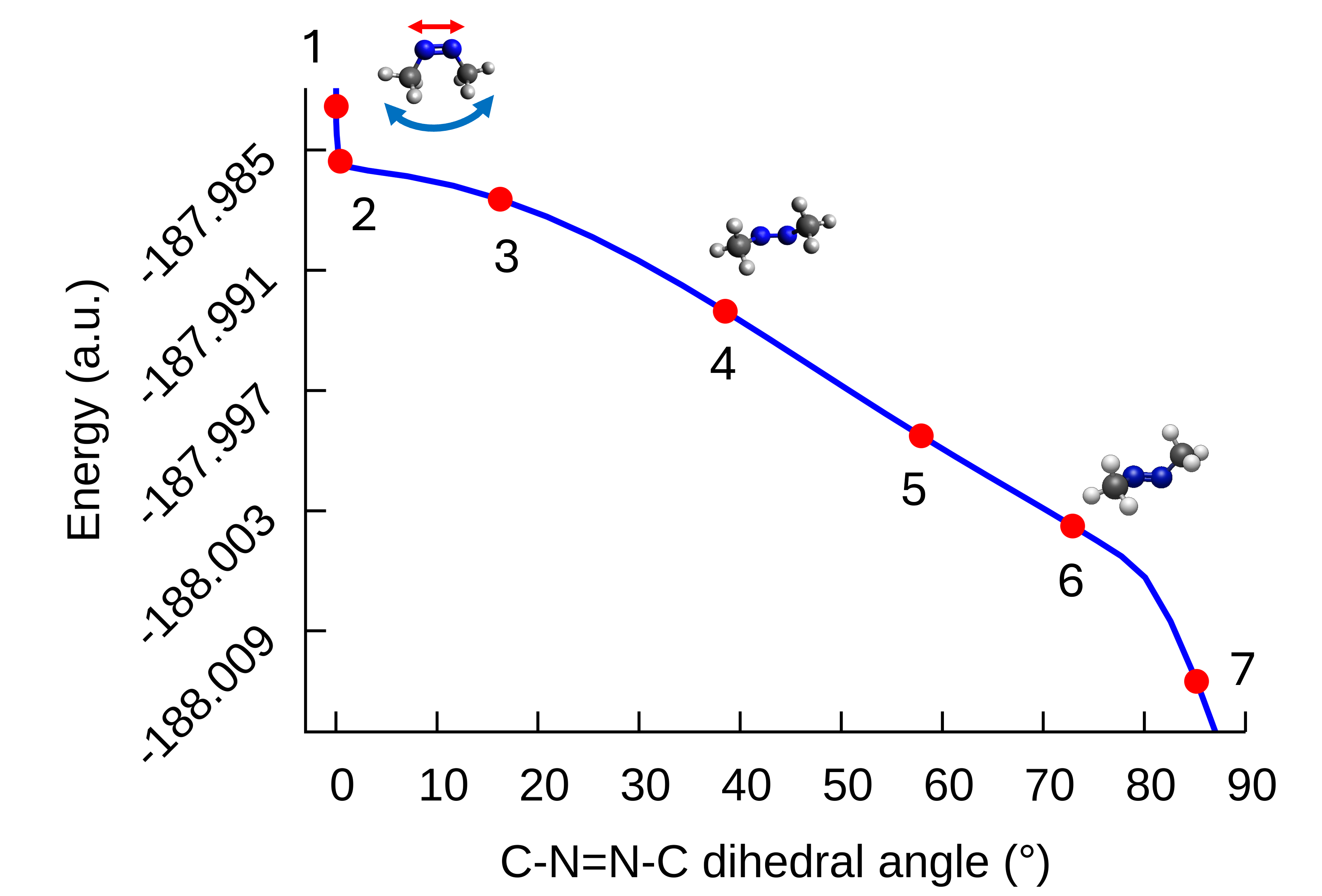}
\\
\includegraphics[width=8.6cm]{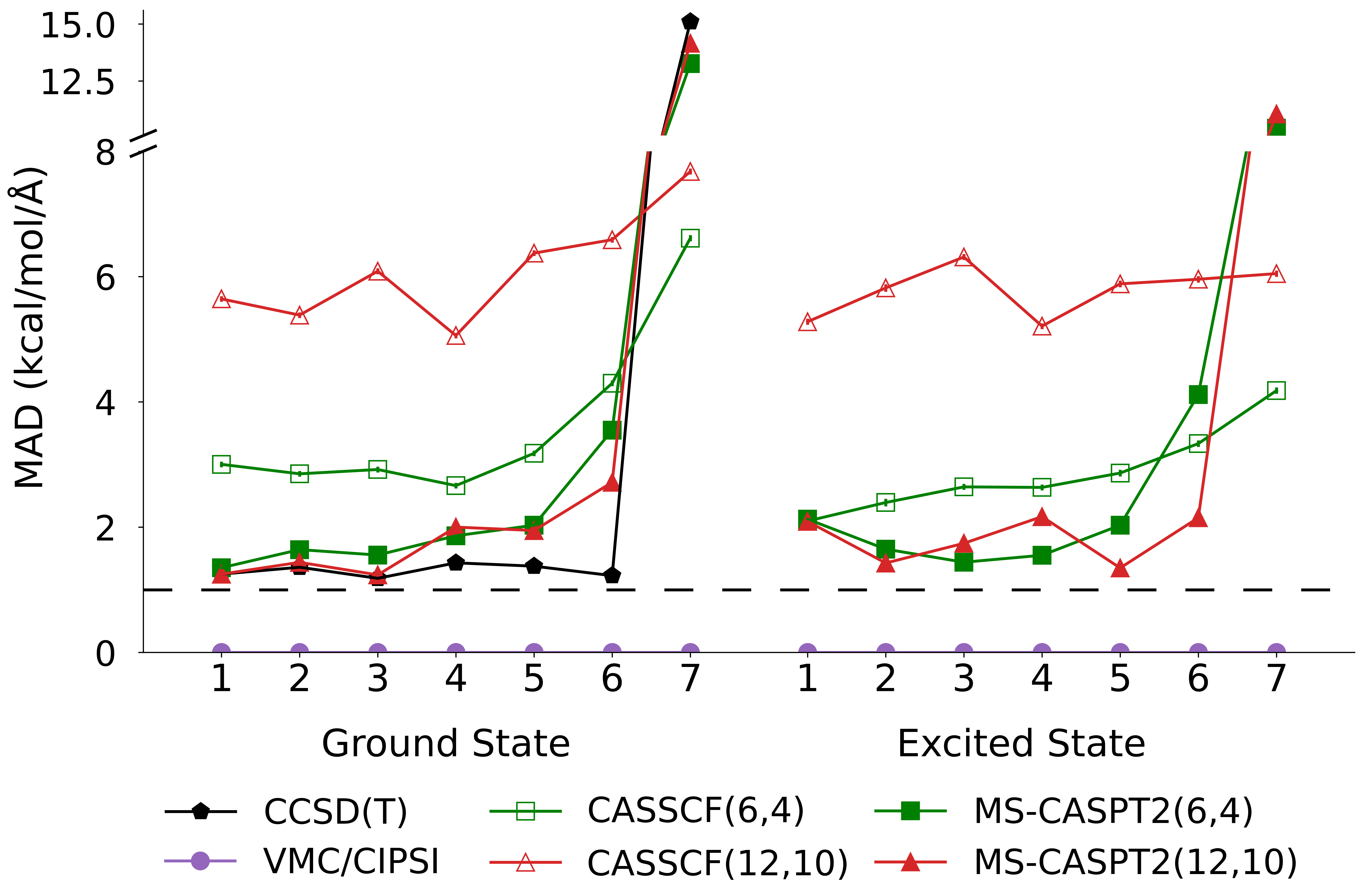}
\caption{Force comparison along the \textit{cis} isomerization pathway. 
Top: excited-state CASSCF(6,4) minimum energy path from the \textit{cis} geometry to the conical intersection, with the seven selected configurations marked. Bottom: force mean absolute deviation (kcal/mol/\AA) relative to VMC/CIPSI, for the ground (left) and the first excited state (right). The dashed line marks 1 kcal/mol/\AA,  the typical residual difference due to basis-set and Jastrow-factor choices. CC calculations use cc-pVQZ and CAS-based methods use aug-cc-pVTZ.
}
\label{fig: testing forces mep}
\end{center}
\end{figure}

Figure~\ref{fig: testing forces mep} reports the mean absolute deviations (MADs) of the ground- and excited-state forces relative to VMC/CIPSI for the seven geometries along the CASSCF(6,4) excited-state minimum energy path.
The VMC forces are robustly converged with respect to the CIPSI expansion (see SI), ensuring their accuracy and making the following method comparisons meaningful.
For the ground state, CCSD(T) and VMC/CIPSI forces agree well along most of the path, with deviations of 1.2--1.4 kcal/mol/\AA\ comparable to the residual differences associated with basis-set and Jastrow-factor choices. This agreement breaks down at configuration 7, close to the conical intersection, where the increasing multireference character makes coupled-cluster methods less reliable for ground-state forces~\cite{cc_review}. Consistently, the  $T_1$ diagnostic at configuration 7 is about 0.06, above the usual closed-shell warning value of 0.02 \cite{t1_diagnostic}.

Turning to the active-space methods, CASSCF(6,4) differs substantially from VMC/CIPSI for both the ground and first excited states, with errors increasing as the conical intersection is approached. Enlarging the active space to (12,10) worsens the agreement, with deviations reaching 7.5 kcal/mol/\AA. The inclusion of the perturbative correction through MS-CASPT2 brings the forces closer to VMC/CIPSI for most configurations and for both active spaces. However, for configuration 7, MS-CASPT2 shows discrepancies larger than 10 kcal/mol/\AA\ in both electronic states, consistent with its known difficulties near degeneracies \cite{Granovsky_xms}. XMS-CASPT2 results are similar to MS-CASPT2 and only partly improve the description near the conical intersection, reducing the deviations to below 8 kcal/mol/\AA\ for both states (see Figure~S7).

\begin{figure}[htbp]
\begin{center}
\includegraphics[width=6.4cm]{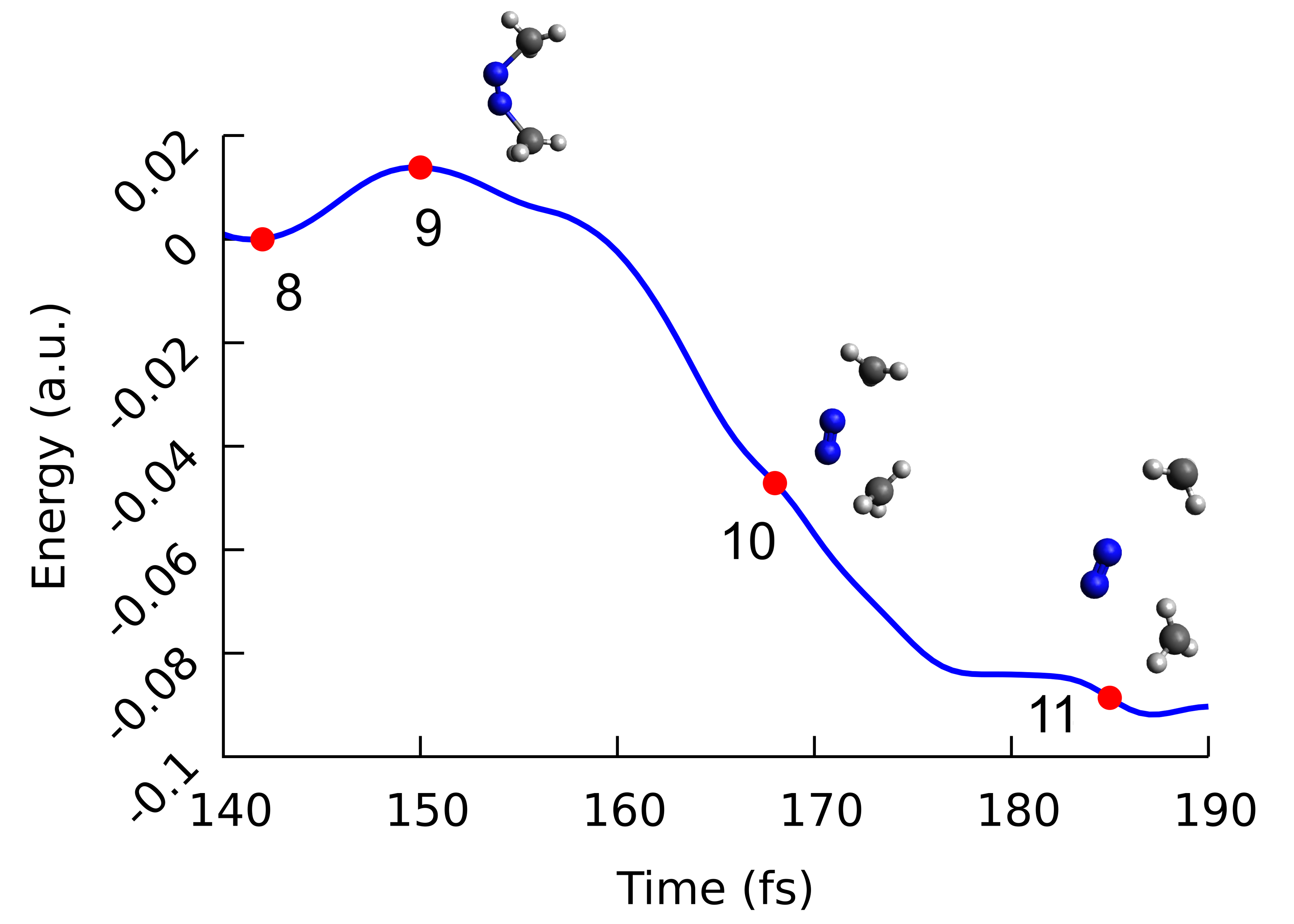}
\\
\includegraphics[width=8.6cm]{Azomethane/Figures/MAD_forces_sharc1210.png}
\caption{Force comparison along a C--N dissociation pathway. Top: ground-state trajectory from a CASSCF(12,10) surface-hopping simulation showing C--N bond cleavage, with the four selected configurations marked as 8--11. Bottom: force mean absolute deviation (kcal/mol/\AA) relative to VMC/CIPSI. The dashed line marks 1 kcal/mol/\AA. CC calculations use cc-pVQZ and CAS-based methods use aug-cc-pVTZ.}
\label{fig: testing forces sharc1210}
\end{center}
\end{figure}

We next consider geometries extracted from a CASSCF(12,10) surface-hopping trajectory undergoing ground-state C--N dissociation, as shown in Figure~\ref{fig: testing forces sharc1210}. In this region, CCSD(T) ground-state forces remain close to VMC as long as the molecule is bound. As dissociation proceeds, the increasing multireference character leads to a breakdown of coupled-cluster accuracy, with the MAD of the forces rising from 1.4 to 5 kcal/mol/\AA. CASSCF(6,4) ground-state forces also deteriorate as the C--N bond is stretched, likely because the active space does not include the relevant $\sigma$ and $\sigma^*$ orbitals. This limitation is reduced in CASSCF(12,10), where these orbitals are included and the force deviations vary less strongly along the dissociation pathway.  MS-CASPT2(12,10) improves the agreement further, keeping the ground-state force deviations within 4 kcal/mol/\AA.

For the excited state, the force deviations remain moderate for the bound geometries, with MS-CASPT2(12,10) giving the closest agreement with VMC/CIPSI. For dissociated geometries, the accuracy drops significantly, with the MAD on the forces exceeding 20 kcal/mol/\AA. 
This large error is due to both active spaces yielding an incorrect dominant $\pi \rightarrow \pi^*$ excitation localized on the \ce{N2} fragment, while the CIPSI expansion correctly captures a lower-energy excitation involving orbitals localized on the \ce{CH3} radical.
Since C--N dissociation is expected to occur only after internal conversion to the ground state through the conical intersection
\cite{Diau_concert,Diau_femtochem}, these excited-state force errors should not directly affect the dissociation dynamics. Their impact is limited to the state-averaged orbitals entering the ground-state description.
Only in the rare CASSCF(12,10) trajectories that dissociate on the excited state can the incorrect excited-state character affect the subsequent fragment dynamics and, to a lesser extent, the estimated excited-state population.

The single-dissociation case is discussed in the SI. It shows a similar breakdown of coupled cluster for the ground state. Even though CASSCF(12,10) correctly recovers the dominant excitation on the $\ce{CH3N2^{\bullet}}$ fragment, the forces still differ significantly from VMC/CIPSI also in the excited state.

Taken together, these results show that restricting the wave function to a fixed active space can lead to a geometry-dependent accuracy of the forces along both the relaxation and C--N dissociation pathways. This limitation also persists for Jastrow-Slater wave functions, as shown by the VMC/CAS(6,4) results reported in the SI. 
In contrast, targeting a fixed PT2 energy correction in the CIPSI expansions leads to a more consistent accuracy for VMC/CIPSI across the different geometries. Indeed, the number of determinants in the VMC/CIPSI wave functions adapts strongly to the geometry, ranging from 748 to 2856 along the minimum energy path and decreasing to as few as 149 for dissociated geometries. Finally, the convergence test over the PT2 threshold ensures that the chosen value yields highly accurate forces.

\subsection{Nonadiabatic dynamics from \textit{cis}-azomethane}

\begin{figure*}[tbhp]
\includegraphics[height=4.5cm]{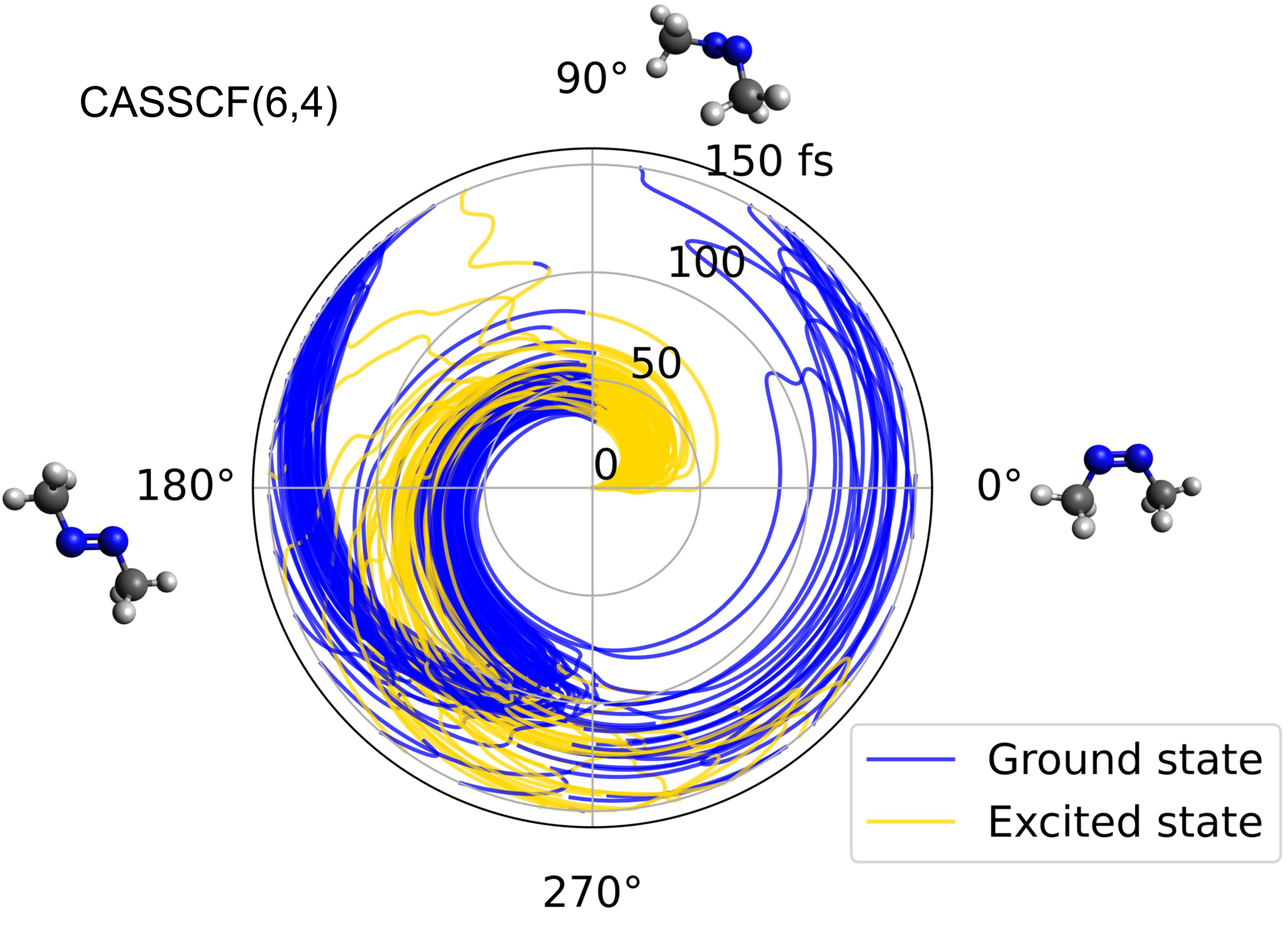}
\includegraphics[height=4.5cm]{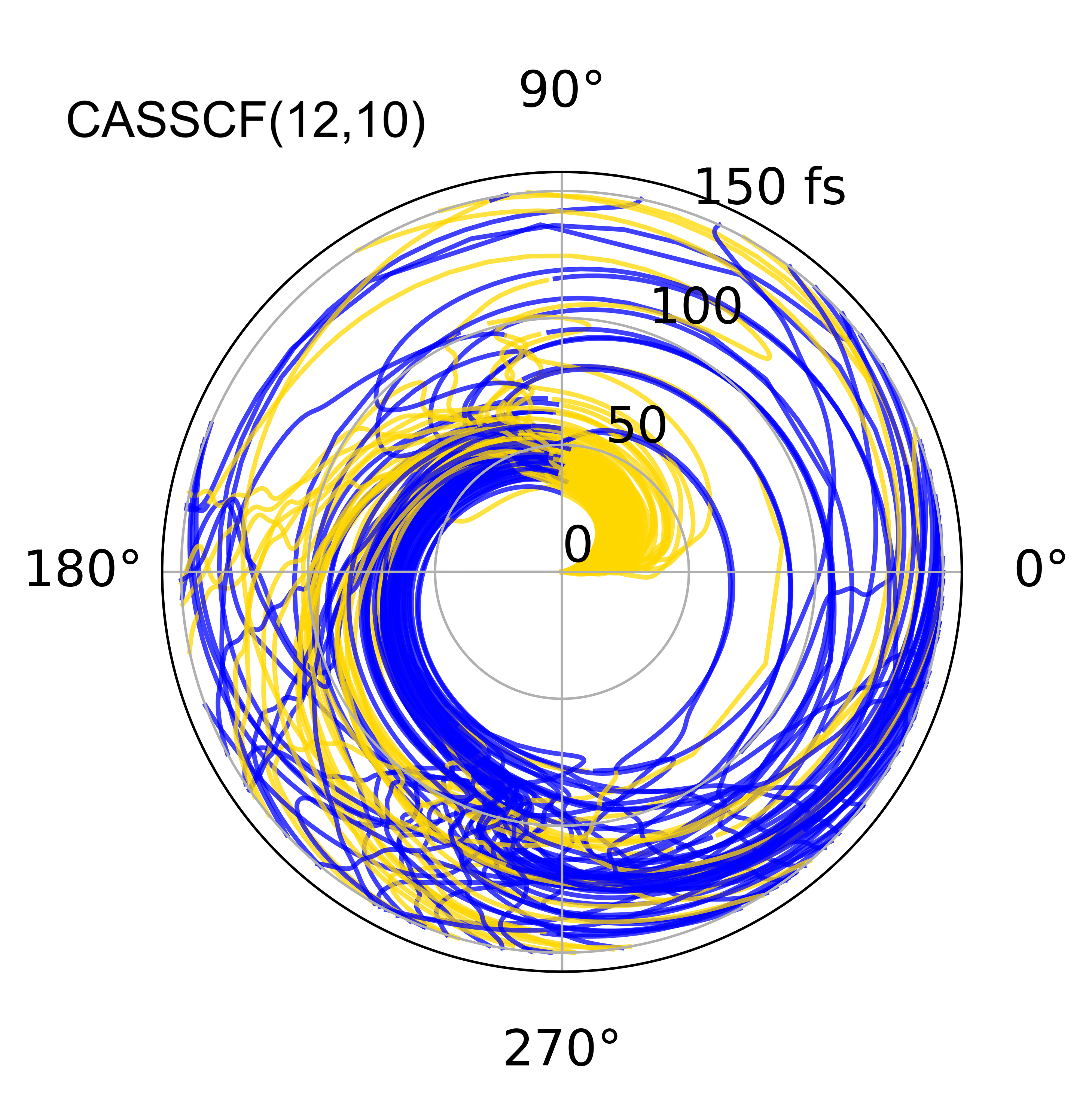}
\includegraphics[height=4.5cm]{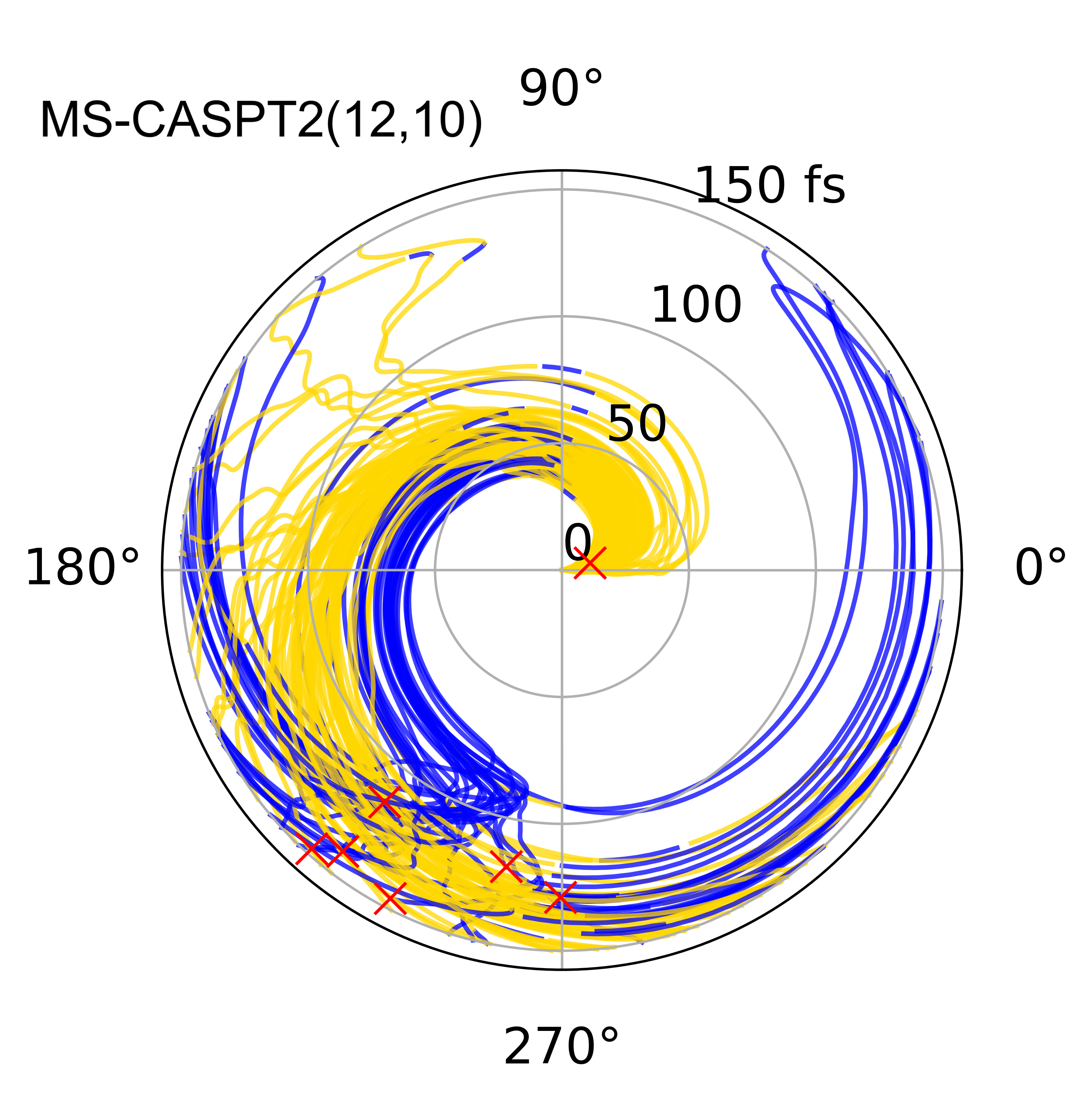}
\caption{Polar representation of 100 $\kappa$TSH \textit{ab initio} trajectories for CASSCF(6,4) (left), CASSCF(12,10) (middle), and MS-CASPT2(12,10) (right). The angular coordinate is the $\angle\text{CNNC}$ dihedral angle and the radius is the elapsed simulation time. Colors indicate the active state, with ground state in blue and excited state in yellow. Red crosses mark crashed trajectories.}
\label{fig: traj polar plot ktdc}
\end{figure*}

We first examine how the force differences identified above manifest in \textit{ab initio} dynamics. Starting from the excited state of the \textit{cis} isomer, 100 $\kappa$TSH trajectories are propagated for 150 fs using CASSCF(6,4), CASSCF(12,10), and MS-CASPT2(12,10). To reduce the cost of the dynamics, these calculations use the aug-cc-pVDZ basis set (basis-set effects are reported in the SI). Figure~\ref{fig: traj polar plot ktdc} shows the evolution of the $\angle$CNNC dihedral angle, the key coordinate for azomethane isomerization, for all trajectories. In these polar plots, the dihedral angle defines the angular coordinate and the elapsed time the radial coordinate.

The CASSCF(6,4) trajectories capture torsion around the N=N bond and yield two decay pathways: ``normal'' trajectories, which relax to the ground state via the first conical intersection at roughly 90$^{\circ}$, and ``rotator'' trajectories, which reach a symmetry-equivalent intersection after an overall rotation of about 270$^{\circ}$ \cite{Vacher_approx,Sellner_photodyn}. 
As expected, CASSCF(6,4) does not produce any C–N dissociation because the active space lacks the corresponding bonding and antibonding $\sigma$ orbitals. With CASSCF(12,10), these orbitals are included and C--N dissociation occurs frequently, with 11\% of the trajectories dissociating already within 40 fs. After bond breaking, the fragments rotate freely, leading to the broad spread of curves in the CASSCF(12,10) polar plot.

This early dissociation is overestimated at the CASSCF level. Higher-level calculations~\cite{Sellner_photodyn,Hennefarth_lpdft} find less dissociation over similar timescales and no C--N bond breaking before the molecule reaches the first conical intersection, namely, within the first few tens of femtoseconds.
Including the PT2 correction yields more realistic dissociation dynamics, with the first bond breaking events occurring only after 94 fs, while preserving the N=N torsional mechanism.
The MS-CASPT2 trajectories are however more prone to convergence failures, particularly in bond-breaking regions, likely because of active-space instabilities as orbitals rotate in and out of the active space \cite{Barbatti_fullycorrelated}.




These short \textit{ab initio} trajectories establish the principal decay pathways and show that C–N bond cleavage depends strongly on the electronic structure method. To make QMC-based nonadiabatic dynamics feasible, we train a multi-state ML force field on the VMC/CIPSI reference data. For direct comparison, analogous CASSCF(12,10) and MS-CASPT2(12,10) models are trained on the same set of configurations. Each model is used to propagate 1000 surface-hopping trajectories up to 400 fs under identical initial conditions.  While much longer timescales, on the order of 100 ps, would be needed to observe near-complete dissociation across the ensemble \cite{cattaneo_persico_azoinsolution}, these lie beyond the extrapolation regime of the ML models and are not pursued here. 


\begin{figure}[htbp]
\begin{center}
\includegraphics[width=8.5cm]{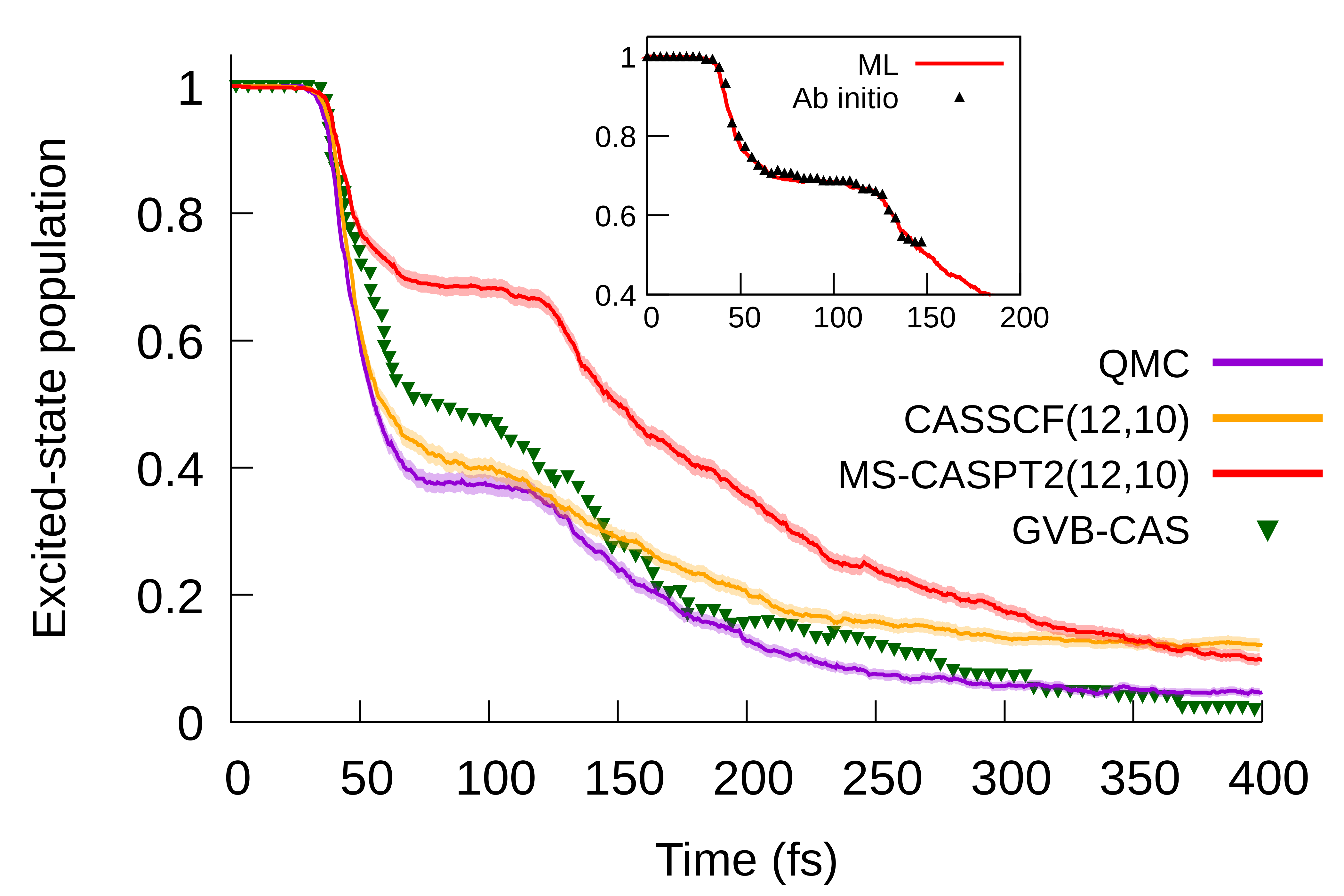}
\caption{Excited-state population as a function of time for the \textit{cis}-azomethane dynamics obtained with the different ML models. Shaded regions indicate the statistical uncertainty estimated from 1000 trajectories. Generalized-valence-bond (GVB)-CAS data are taken from Ref.~\citenum{Sellner_photodyn}.
The inset compares the reference \textit{ab initio} MS-CASPT2 population with the corresponding ML prediction. }
\label{fig: state population cis vmc}
\end{center}
\end{figure}


For these larger ML ensembles, we focus on excited-state populations, pathway statistics, and product yields. Figure~\ref{fig: state population cis vmc} shows the time evolution of the excited-state population. All methods exhibit an initial latency of about 35 fs, followed by rapid population transfer to the ground state within the first 60 fs as trajectories reach the first conical intersection. A residual excited-state population then persists for roughly 50 fs, corresponding to trajectories that continue toward the \textit{trans} region and partially return to \textit{cis}. These trajectories decay when they encounter the second, symmetry-related conical intersection. As a check on the ML representation, the inset shows the close agreement between the MS-CASPT2 population from the ML model and the corresponding \textit{ab initio} result. 

The main method-dependent differences appear after the trajectories reach the first conical-intersection region. MS-CASPT2(12,10) predicts a slower population transfer, with more than 70\% of the trajectories remaining on the excited state beyond this point, compared with about 40\% for QMC and CASSCF(12,10). The profile of the QMC population is also broadly compatible with the generalized-valence-bond (GVB)-CAS dynamics of Sellner \textit{et al.}~\cite{Sellner_photodyn}, which lies between the CASSCF and QMC curves over part of the decay. This method augments CAS(6,4) with a GVB perfect-pairing description of the nine single bonds. 

The difference between the MS-CASPT2 population and the QMC and CASSCF populations reflects the fact that, while two thirds of the MS-CASPT2 trajectories follow the rotator pathway remaining in the excited-state longer, only one third of the QMC and CASSCF trajectories continue on this pathway (see Table~\ref{tab: traj lifetimes}). We note that the normal/rotator classification excludes trajectories that do not return to the ground state within 400 fs (namely, 3.5\% of the QMC and 7.5\% of both the CASSCF and MS-CASPT2 trajectories).  The slower MS-CASPT2 decay relative to CASSCF is consistent with previous studies~\cite{Papineau,Xu_ultrafast} and develops in the part of the dynamics where the description of the conical-intersection region is most critical. This is also where MS-CASPT2 shows the largest deviations from VMC/CIPSI as illustrated by configuration 7 in Figure~\ref{fig: testing forces mep}. The XMS-CASPT2 variant gives a qualitatively similar slow population decay as reported in the SI.

To understand the difference between the QMC and CASSCF populations at later times, one needs to separate the effect of normal/rotator pathway branching, which is similar for the two methods, from the lifetimes within each channel. To this end, the excited-state populations are fitted separately for the normal and rotator trajectories  according to $\min \left(1,\exp \left[-(t-\tau_1)/\tau_2 \right]\right)$, and the corresponding latency ($\tau_1$) and decay ($\tau_2$) times are reported in Table~\ref{tab: traj lifetimes}. 

QMC, CASSCF, and MS-CASPT2 give similar latency and decay times along the normal pathway but display larger differences for the rotator trajectories. In particular, the longer rotator lifetime obtained with CASSCF explains why the QMC and CASSCF populations agree during the initial decay but diverge at later times. The fit of the GVB-CAS data~\cite{Sellner_photodyn} differs more strongly for the normal pathway but remains qualitatively consistent in terms of the overall excited-state lifetime $\tau_1+\tau_2$. Table~\ref{tab: traj lifetimes} also reports the average hopping torsional angles which are similar for all methods, with values close to $92^\circ$ and $267^\circ$ for the first and second conical-intersection passages, respectively. This confirms that the two decay channels correspond to symmetry-related conical-intersection regions. 



\begin{table*}[t]
\centering
\caption{Latency time ($\tau_1$), decay time ($\tau_2$), average $\angle$CNNC torsional angle at hopping ($\langle \theta \rangle$), and dissociation yield at 400 fs for trajectories initiated from the \textit{cis} and \textit{trans} isomers. For the \textit{cis} photodynamics, the fitted times and hopping angles are reported separately for normal and rotator trajectories.}
\begin{tabular}{lccccc}
\hline \addlinespace
Method & no.\ traj
 & $\tau_1$ (fs)
 & $\tau_2$ (fs)
 & $\langle \theta \rangle$ (deg)
 & \% dissociation \\
\hline \addlinespace
\multicolumn{6}{c}{Starting from \textit{cis} (normal trajectories)}\\
QMC           & 632 & 36.8 & 12.6 & $91.9 \pm 0.3$ & 43.5\\
CASSCF(12,10) & 596 & 37.4 & 13.7 & $93.4 \pm 0.4$  & 95.3\\
MS-CASPT2(12,10) & 315 & 37.4 & 11.1 & $91.8 \pm 0.9$ & 68.3\\
GVB-CAS$^{a}$ & 63 & 29   & 37   & $94 \pm 5$ & 14.3 \\
\multicolumn{6}{c}{Starting from \textit{cis} (rotator trajectories)}\\
QMC           & 333 & 116.6 & 61.4 & $267.2 \pm 0.6$ & 16.2\\
CASSCF(12,10) & 329 & 107.1 & 84.4 & $265 \pm 1$  & 84.2\\
MS-CASPT2(12,10) & 610 & 116.8 & 94.7 & $265.0 \pm 0.6$ & 41.8\\
GVB-CAS$^{a}$ & 36 & 104   & 82   & $266 \pm 8$ & 2.8 \\
\multicolumn{6}{c}{Starting from \textit{trans}}\\
QMC           & 1000 & 77.4 & 77.7 & $95.5 \pm 0.6$ & 9.5\\
CASSCF(12,10) & 1000 & 72.7 & 82.8 &  $95.4 \pm 0.6$ & 82.3 \\
MS-CASPT2(12,10) & 1000 & 66.6 & 121.4 & $93.9 \pm 0.5$ & 29.7\\
GVB-CAS$^{a}$ & 100 & 96   & 89    & $95 \pm 6$ & 0 \\
MRCI(6,4)+PP$^{a}$  & 20 & 85   & 78  & $93 \pm 5$ & 0 \\
FOMO-CI(6,6)$^{b}$  & 5000 & 73 & 86 & & 29.9 \\
\addlinespace \hline
\addlinespace
\multicolumn{6}{p{12cm}}{\small $^{a}$ Initial conditions generated with the Wigner sampling at the same level of theory and without torsional bias \cite{Sellner_photodyn}.} \\
\multicolumn{6}{p{12cm}}{\small $^{b}$ Local diabatization algorithm for propagation \cite{Pieroni_initconds}.} \\
\end{tabular}
\label{tab: traj lifetimes}
\end{table*}

\begin{figure}[htbp]
\begin{center}
\includegraphics[width=8.8cm]{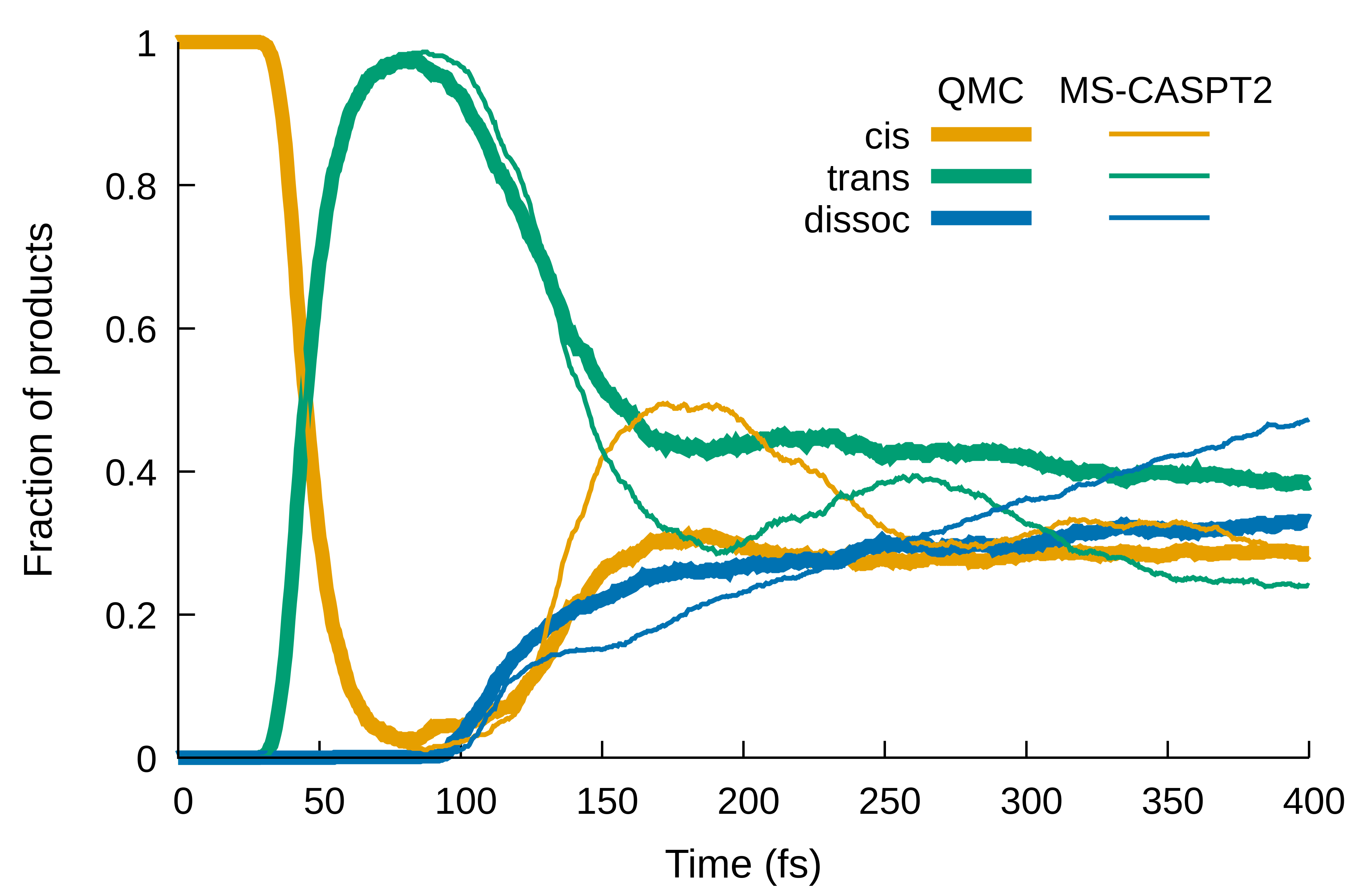}
\caption{Fractions of azomethane trajectories in \textit{cis} ($|\angle \text{CNNC}|<90^{\circ}$), \textit{trans} ($|\angle \text{CNNC}-180^\circ|<90^{\circ}$), and dissociated configurations as a function of time for the \textit{cis}-iniziated dynamics. Single and double dissociations are grouped together and identified when either or both C-N bond lengths exceed 2.25 \AA.}
\label{fig: products vmc}
\end{center}
\end{figure}

We next consider C--N bond cleavage and the resulting product yields. Both QMC and MS-CASPT2 strongly reduce the excessive fragmentation obtained with CASSCF. At 400 fs, the fraction of trajectories with at least one dissociated C--N bond is 33\% for QMC and 47\% for MS-CASPT2, compared with 92\% for CASSCF. Figure~\ref{fig: products vmc} shows the corresponding time-dependent product distribution, where single and double dissociations are grouped together. QMC and MS-CASPT2 give similar yields up to about 120 fs, with most trajectories first converting from the \textit{cis} to the \textit{trans} isomer and then partially returning toward \textit{cis}. This reconversion is more pronounced for MS-CASPT2, consistent with its larger fraction of rotator trajectories, and leads to weak oscillations in the product populations that are absent in QMC. At longer times, dissociation accumulates in both simulations but more rapidly with MS-CASPT2 than with QMC.

Despite the quantitative differences in total dissociation yield, the pathway dependence is consistent across QMC, MS-CASPT2, and the GVB-CAS dynamics of Sellner \textit{et al.}~\cite{Sellner_photodyn}: normal trajectories are more likely to dissociate than rotator trajectories. This trend is also visible in Table~\ref{tab: traj lifetimes}, where the GVB-CAS yields are much smaller in absolute value but follow the same ordering between the two pathways.

\subsection{Nonadiabatic dynamics from \textit{trans}-azomethane}

We next apply the same ML models to nonadiabatic dynamics initiated from \textit{trans}-azomethane. The models are trained on the dataset generated from the \textit{cis}-initiated dynamics and are used here without additional \textit{trans}-specific training. This extension is justified by the fact that the \textit{cis}- and \textit{trans}-initiated trajectories probe the same $\angle\text{CNNC}$ torsional coordinate and the same possible C--N bond-cleavage channels after internal conversion. At the same time, the \textit{trans} dynamics provides a simpler case, with a single passage through the conical-intersection region, and offers a more direct connection to ultrafast photofragmentation experiments.

\begin{figure}[htbp]
\begin{center}
\includegraphics[width=8.5cm]{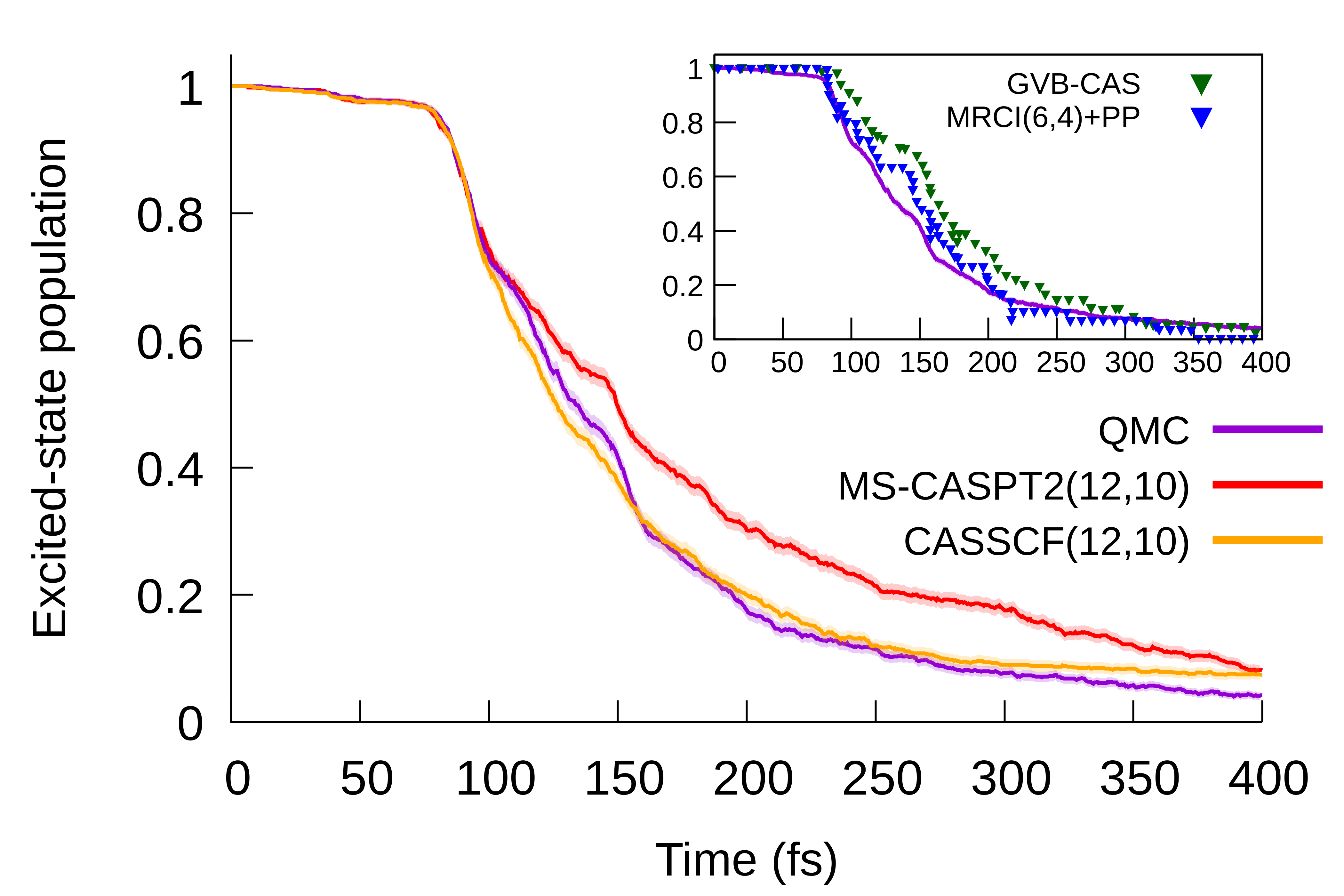}
\caption{Excited-state population as a function of time for the photodynamics of \textit{trans}-azomethane, computed with the different ML models. Shaded regions represent the statistical uncertainty estimated over 1000 trajectories. The inset compares the QMC population with data from Ref.~\citenum{Sellner_photodyn}.
}
\label{fig: state population trans vmc}
\end{center}
\end{figure}

As shown in Figure~\ref{fig: state population trans vmc}, the excited-state population starting from \textit{trans}-azomethane has a longer latency time (about 80 fs), followed by an approximately single-exponential decay. Unlike in the \textit{cis}-initiated dynamics, the trajectories do not cross the conical-intersection region a second time during the rotation. QMC and CASSCF give similar population decays, while MS-CASPT2 again predicts a slower transfer to the ground state, with a larger decay constant (Table~\ref{tab: traj lifetimes}). The QMC population is also close to the GVB-CAS and MRCI results of Ref.~\citenum{Sellner_photodyn} as well as to the floating occupation molecular orbital (FOMO)-CI data of Ref.~\citenum{Pieroni_initconds}.


\begin{figure}[htbp]
\begin{center}
\includegraphics[width=8.1cm]{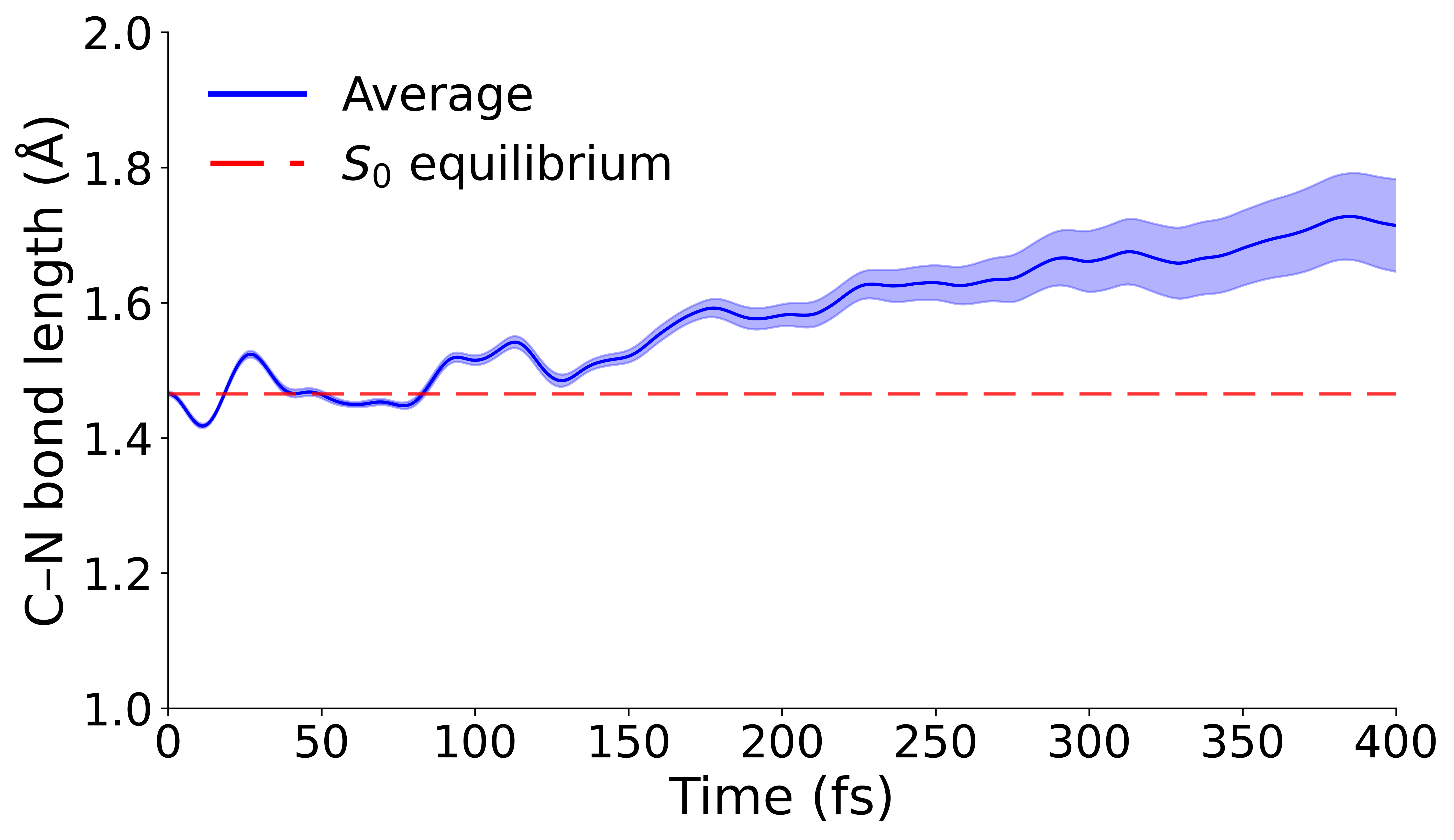}
\\
\includegraphics[width=8.2cm]{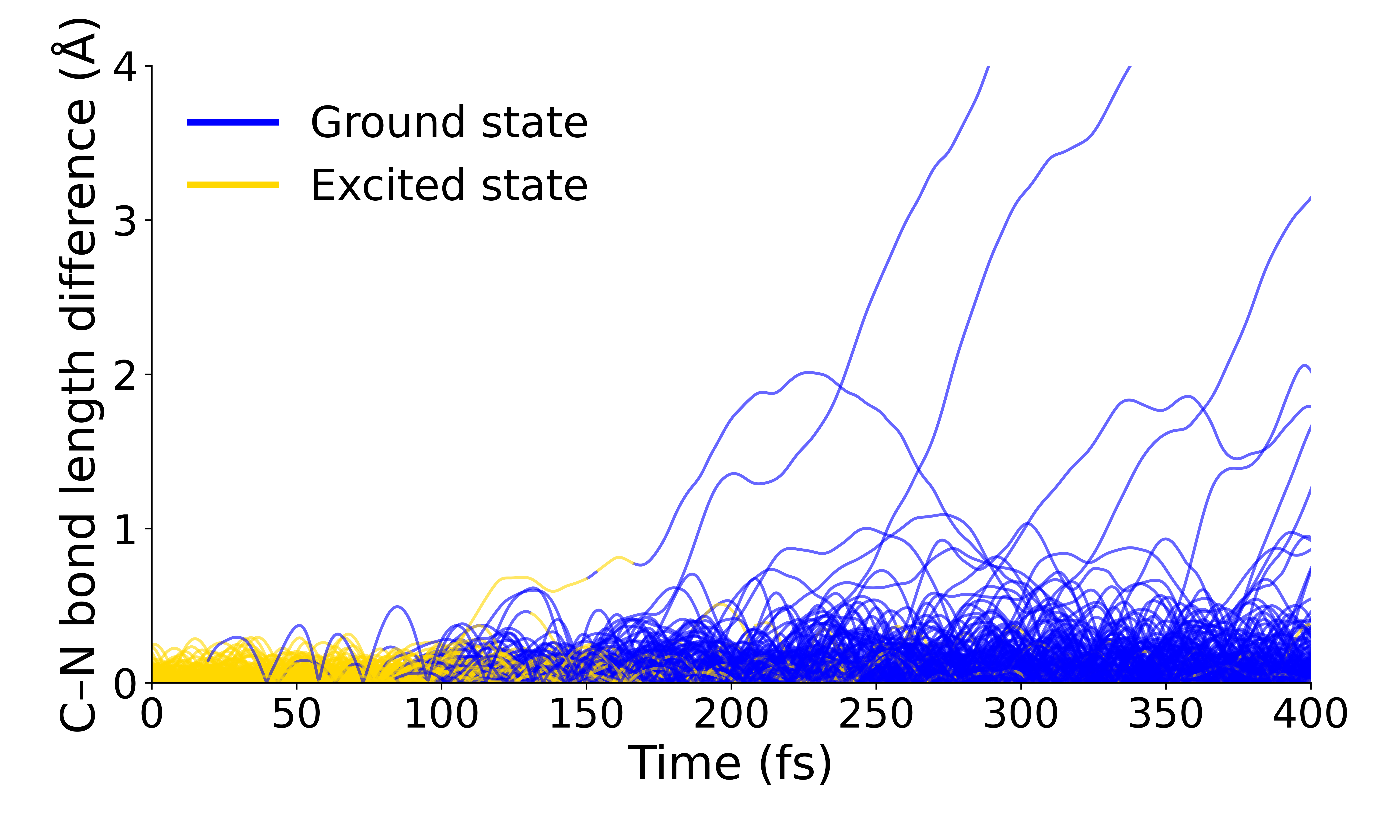}
\caption{C--N bond-length analysis for the QMC-trained \textit{trans}-azomethane dynamics. Top: average C--N bond length over all trajectories, with bond stretching becoming apparent after about $160$ fs. Bottom: absolute difference between the two C--N bond lengths for a subset of 100 trajectories, indicating single versus sequential double C--N dissociation.}
\label{fig: bond length vmc}
\end{center}
\end{figure}

The similarity between QMC and CASSCF in the excited-state population does not extend to the fragmentation dynamics. As in the \textit{cis}-initiated simulations, CASSCF strongly overestimates C--N bond cleavage, giving a dissociation yield of 82\% after 400 fs. MS-CASPT2 reduces this value to 30\%, while QMC predicts a much smaller dissociation yield of 9.5\%. In the QMC dynamics, C--N bond breaking becomes apparent only after about 160 fs, as also reflected in the increase of the average C--N bond length shown in Figure~\ref{fig: bond length vmc}.


To further characterize the fragmentation mechanism, Figure~\ref{fig: bond length vmc} also shows the absolute difference between the two C--N bond lengths for a subset of 100 trajectories. A sustained increase of this difference indicates single C--N bond cleavage, while sequential breaking of both C--N bonds gives an initial increase followed by a plateau or partial decrease. In this subset, 8 trajectories dissociate, with only one undergoing double dissociation. Over the full ensemble of 1000 trajectories, the corresponding yields are 7.2\% for single dissociation and 2.3\% for double dissociation. All fragmentation events occur after passage through the conical intersection and no QMC trajectory dissociates on the excited state. Additional details on the time-dependent product distribution are provided in Section S10.

The QMC dynamics therefore indicate a small but non-negligible C--N dissociation component after internal conversion, with bond breaking starting around 160 fs and accumulating over the 400 fs simulation time. This differs from the GVB-CAS trajectories, where C--N bond breaking was observed only when the initial conditions were explicitly biased toward N=N torsion~\cite{Sellner_photodyn}. The onset is instead similar to the more recent FOMO-CI dynamics of Persico and co-workers, which also reported a reaction onset around 160 fs, although with a larger dissociation yield of 29.9\% after 400 fs~\cite{Pieroni_initconds}.

The comparison with earlier simulations and experiments mainly concerns the onset time. In the experiments of Diau and Zewail, \ce{CH3N2} fragments were observed within 70--100 fs~\cite{Diau_femtochem,Diau_concert}.  This indicates that at least part of the dissociation occurs on an ultrafast timescale, although the experimental onset is earlier than in the QMC trajectories. As argued by Sellner \textit{et al.}~\cite{Sellner_photodyn}, partial vibrational excitation of the torsional mode under the experimental conditions could contribute to this earlier onset. Earlier simulations by Cattaneo and Persico, based on trajectory surface hopping on PESs fitted to CIPSI energies with zero-point-energy corrections and classical Boltzmann initial sampling, instead found \ce{CH3N2} formation only at approximately 250 fs~\cite{cattaneo_persico_azoinsolution}.

Overall, these results support an impulsive component in which a small fraction of azomethane trajectories dissociates shortly after internal conversion to the ground state, while the majority is expected to fragment on a longer, statistical timescale, of the order of hundreds of picoseconds~\cite{cattaneo_persico_azoinsolution}. Quantitative comparisons of both onset times and dissociation yields should, however, be made cautiously across studies. They depend on the bond-length threshold used to define C--N cleavage, set here to 2.25~\AA, and on the initial sampling strategy. In particular, Wigner sampling includes zero-point vibrational energy, which can leak into dissociative coordinates in classical dynamics and accelerate dissociation relative to classical Boltzmann sampling~\cite{Pieroni_initconds,Uzer_zpe,Barbatti_zpe}. 

\section{Conclusions}

We have introduced multi-state machine-learned nonadiabatic dynamics based on VMC/CIPSI reference energies and forces. The underlying QMC wave functions combine compact selected determinantal expansions with an explicit treatment of dynamical correlation. Neural-network force fields provide a smooth representation of the stochastic data and make it possible to propagate large surface-hopping ensembles without direct on-the-fly QMC dynamics. 

For azomethane, the QMC reference data are robustly converged across torsional relaxation and C–N dissociation geometries. The ground-state forces agree closely with CCSD(T) where a single-reference description is reliable. Near the conical-intersection region and in dissociated geometries, fixed-active-space methods show larger and more geometry-dependent deviations, while the QMC protocol maintains a consistent description near the conical intersection and along bond-breaking pathways where the electronic character changes significantly.

The ML-based dynamics confirms the strong sensitivity of azomethane photochemistry to the underlying electronic structure method. CASSCF substantially overestimates C--N bond breaking, while the VMC/CIPSI-trained dynamics suppresses this excessive dissociation, preserving the expected torsional relaxation through the conical-intersection region. CASPT2 also reduces the CASSCF fragmentation but predicts slower excited-state population transfer and larger dissociation yields than QMC. The lower QMC dissociation yield does however not imply the absence of early fragmentation: after internal conversion to the ground state, a small but non-negligible fraction of trajectories undergoes C--N bond breaking within the first few hundred femtoseconds.

Starting from the \textit{trans} isomer, QMC again predicts much less dissociation than CASSCF and CASPT2, with fragmentation occurring only after passage through the conical-intersection region. The bond-length analysis shows that single C--N bond cleavage is more frequent than sequential double dissociation within the simulated time window. The QMC dynamics therefore supports a small but non-negligible prompt dissociation component after internal conversion, in qualitative agreement with femtosecond-resolved experiments, although quantitative comparisons of onset times and yields depend on the bond-breaking criterion and on the initial sampling strategy.

These results show that multi-state QMC-trained ML force fields can bring correlated wave-function accuracy to large-ensemble nonadiabatic photodynamics, spanning regions of configuration space that differ significantly in geometry and electronic character.


\vspace*{20px}
\section*{Associated content}
Supporting Information: Basis set convergence and pseudopotential error on vertical excitation energies; basis set and Jastrow factor impact on forces; VMC forces using CAS expansions; convergence of VMC energies and forces with CIPSI expansion size; comparison of MS- and XMS-CASPT2 forces and dynamics; force benchmarks on single-dissociated geometry; neural network hyperparameters and adaptive sampling details; influence of time-derivative coupling approximations on population transfer; evolution of gradient norms along the trajectories; and additional VMC-ML trajectory analysis

\section*{Acknowledgments}
The authors thank Remco Havenith and Thijs Mulder for discussions on the electronic structure and dynamics of azomethane, Anthony Scemama for assistance with Quantum Package, Igor Poltavskyi for suggestions on active learning, and Gerrit-Jan Linker for comments on the manuscript.
This work was supported by the Dutch Research Council (NWO) under grant number OCENW.M.22.239. 
The calculations were performed on the Dutch national supercomputer Snellius with the support of the SURF Cooperative (grant number NWO-2025.003).


\bibliographystyle{achemso}
\bibliography{bibliography}

\end{document}